# Field Target Allocation and Routing Algorithms for Starbugs

Michael Goodwin*[a], Nuria P. F. Lorente[a], Christophe Satorre[b], Sungwook E. Hong[c], Kyler Kuehn[a], Jon S. Lawrence[a]

[a]Australian Astronomical Observatory, PO Box 915, North Ryde, NSW 1670, Australia;
[b]Laboratoire de systèmes robotiques (LSRO), EPFL-STI-IMT-LSRO, ME A3 484 (Bâtiment ME), Station 9, CH–1015 Lausanne, Switzerland; [c]Korea Institute for Advanced Study, 85 Hoegiro, Dongdaemun-gu, Seoul 130-722, Republic of Korea

**ABSTRACT**

Starbugs are miniaturised robotic devices that position optical fibres over a telescope's focal plane in parallel operation for high multiplex spectroscopic surveys. The key advantage of the Starbug positioning system is its potential to configure fields of hundreds of targets in a few minutes, consistent with typical detector readout times. Starbugs have been selected as the positioning technology for the TAIPAN (Transforming Astronomical Imaging surveys through Polychromatic Analysis of Nebulae) instrument, a prototype for MANIFEST (Many Instrument Fiber System) on the GMT (Giant Magellan Telescope). TAIPAN consists of a 150-fibre Starbug positioner accessing the 6 degree field-of-view of the AAO's UK Schmidt Telescope at Siding Spring Observatory. For TAIPAN, it is important to optimise the target allocation and routing algorithms to provide the fastest configurations times. We present details of the algorithms and results of the simulated performance.

## 1. INTRODUCTION

In this paper we document the initial efforts on the field configuration algorithm (FCA) for the new robotic positioner called "Starbugs" [1], in preparation for the planned TAIPAN (Transforming Astronomical Imaging surveys through Polychromatic Analysis of Nebulae) [2] and MANIFEST (Many Instrument Fiber System) [3,4,5] instruments. The FCA involves two key software sub-modules: allocation and routing, with optimisation to provide the fastest field configuration times that also do not significantly imprint artificial structure on observed target distributions. Section 1 outlines the background and the key characteristics of Starbugs and instrumental applications. Section 2 summarises the goals of the allocation and routing software and data model. Section 3 reports on the Allocation Module. Section 4 reports on the Routing Module. Section 5 discusses the Simulator and Results. Concluding remarks are provided in Section 6.

### 1.1 Multi-object Fibre Spectroscopy

Multi-object fibre spectroscopy (MOS) is an established and effective technique in the field of Astronomy [23]. Galaxy surveys have produced hundreds of thousands of spectra enabling a wide range of galaxy properties to be studied, including stellar ages, star formation rates and metallicities. The goal in MOS is to precisely place many optical fibres on specific objects within the telescope's field of view. The light collected from these fibres is then relayed to stable bench-mounted spectrographs. MOS surveys facilitated by the AAT 2dF instrument with the spectrograph include 2dFGRS [6], 2QZ [7], WiggleZ [8], GAMA [9], SAMI [10] and the recently commissioned HERMES instrument: GALAH [11].

Modern surveys have significantly increased the demand for a larger number of objects per field, with flexibility in payload configurations and instrument feeds. The future generation of Extremely Large Telescopes (ELTs) will bring about further challenges for efficient reconfiguration of focal planes for multi-object astronomy. These issues are the motivation for developing the new Starbug versatile robotic fibre positioners.

---

* michael.goodwin@aao.gov.au; http://www.aao.gov.au; phone +61 2 9372 4851; fax +61 2 9372 4860

## 1.2 Robotic Fibre Positioners

The scientific motivation for large-scale redshift surveys has driven the development of efficient robotic fibre positioners to accurately and quickly place fibres on the focal plane. The Australian Astronomical Observatory (AAO) has successfully developed 'pick and place' robotic arm systems, such as 2dF [12] (approx. 400 fibres, see Figure 1) on the Anglo-Australian Telescope (AAT) and OzPoz [13] on the Very Large Telescope. These systems position fibres sequentially with the result that the time taken for a field reconfiguration increases linearly as the number of fibres increases. The 2dF instrument takes up to an hour to reconfigure a typical field. To minimise the downtime during re-configuration, a field plate rotation/exchange system is utilised. The expense of fibre cable duplication and increase in total configuration time becomes inefficient for large telescopes with a high multiplicity of targets. These systems are also somewhat inflexible in positioning certain types of fibre payloads over a spherical surface.

Existing fibre positioning technologies exhibit major deficiencies when appleided to large focal planes such as that expected for ELTs. Some of these challenges include a non-planar focal plane (e.g. spherical curvature with GMT) as well as diverse payload requirements (e.g. single fibres, image slices and mixed format IFUs). The AAO has developed Starbugs to overcome the limitations of existing multi-object robotic fibre positioners.

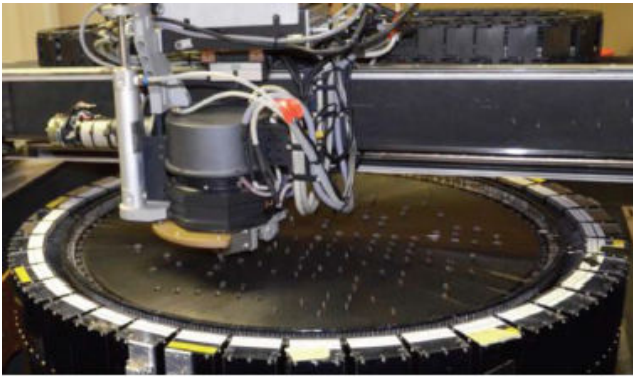 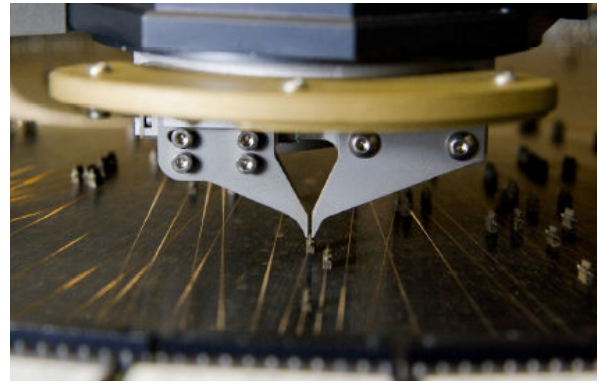

Fibres secured on the field plate of the 2dF instrument        2dF robotic gripper to place fibres on the focal plane

Figure 1: The 2dF robotic positioner for the AAT placing fibres sequentially.

## 1.3 Starbugs

Starbugs are miniature piezoelectric robots that can quickly and accurately position many optical 'payloads' simultaneously. The goal of Starbugs is to reconfigure typical fields in a few minutes over flat or curved focal surfaces. They were first described in 2004 [14], and later in 2006 [15, 16], 2010 [17], 2012 [18] and recently 2014 [1]. An individual Starbug comprises two piezoceramic tube actuators, joined at one end to form a pair of concentric 'legs' that can be electrically driven to produce a micro-stepping motion in the ±x and ±y directions (i.e. forwards, backwards, left, right) and θ (rotation), see Figure 2 (a). The piezoceramic tubes are characterised with an automated LabVIEW application [24]. Discrete step sizes of only a few microns mean that Starbugs are precise, yet they can move several millimetres per second when stepping at high frequencies. Starbugs are positioned under closed-loop control and are monitored with a camera imaging their back-illuminated metrology fibres. The outer diameter of a Starbug is approximately 8mm. The movement characteristics of the Starbugs are listed in Table 1.

Each Starbug carries an individual payload such as an optical fibre, fibre bundle or microlens assembly, which are mounted in the Starbug's central aperture (as shown Figure 2 (b)). The Starbug is held to the glass field plate via a vacuum between the piezoceramic tubes (Figure 3). Starbugs provide a multitude of additional advantages over existing multi-object positioners and therefore two instruments have been proposed: TAIPAN [2] and MANIFEST [3,4,5].

Table 1: Starbug movement specifications for algorithm development

|  | **Precise Positioning** | **Coarse Positioning** | **Rotation** |
|---|---|---|---|
| **Direction** | +/- x, +/- y | +/- x, +/- y | +/- θ |
| **Step Size** | 3 microns | 20 microns | 1 degrees |
| **Speed** | 0.3 mm/s | 2 mm/s | 35 degrees /s |

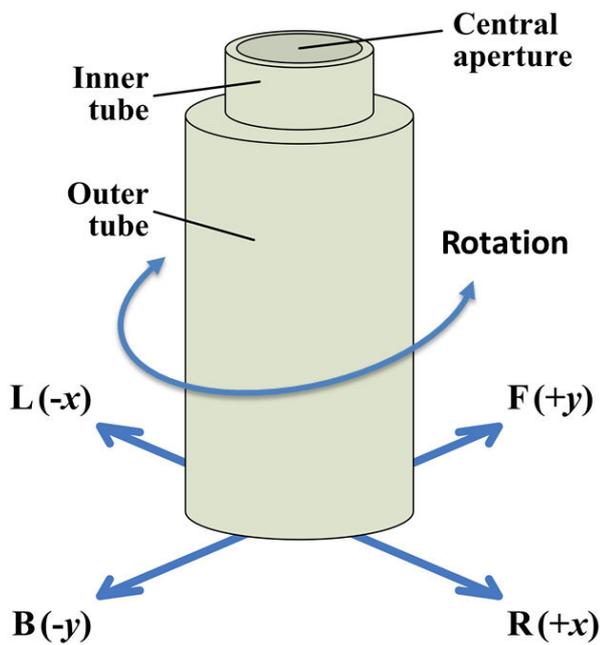

(a) Exterior diagram of the Starbug showing the possible movement directions (forward, right, back, left and rotation).

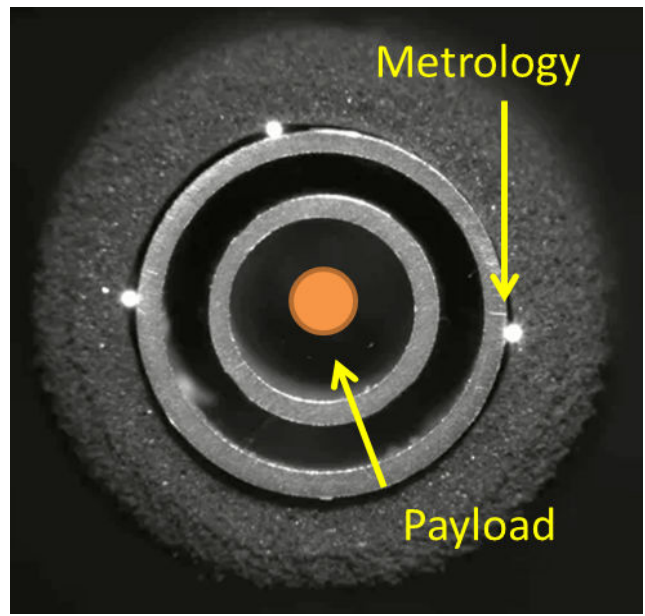

(b) Starbug front surface (contacts with glass field plate to form a vacuum) showing the inner and outer tubes, the metrology fibres and the location of the science payload. A protective jacket is placed on the outer tube during testing.

Figure 2: Starbug mechanical movements (a) and nominal payload position (b).

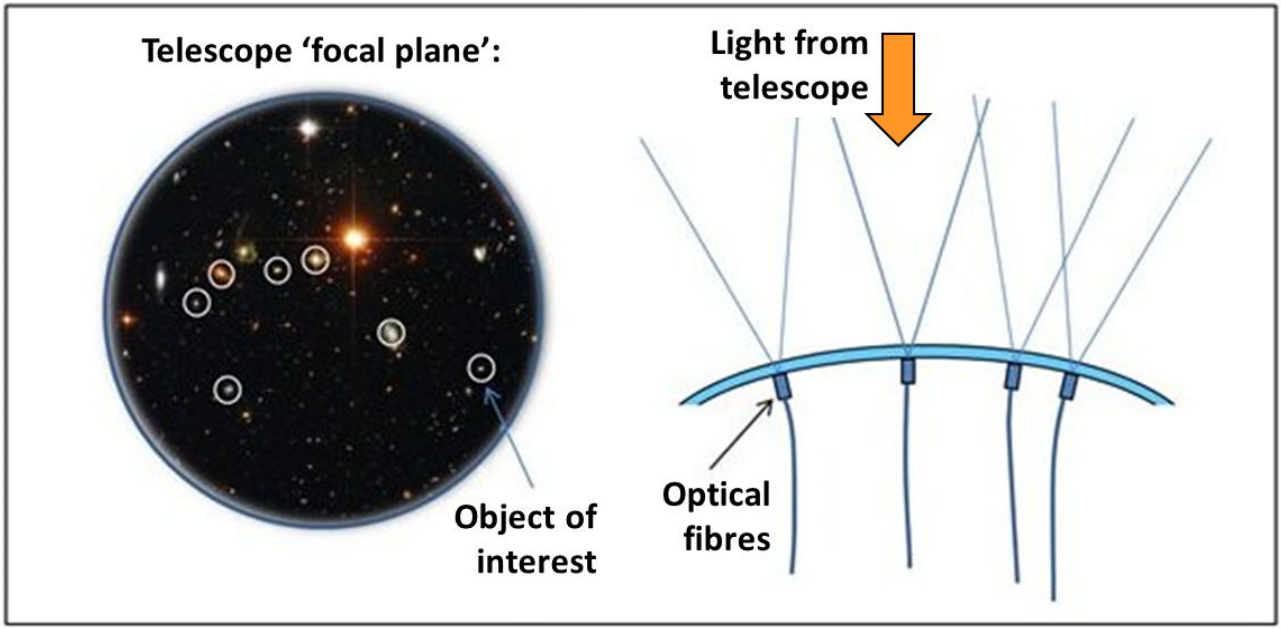

Figure 3: Operational concept showing how the Starbug 'catches' the light from an object on the focal plane (flat or curved) held by a vacuum.

**1.4 Planned Instruments: TAIPAN and MANIFEST**

Starbugs are proposed by the AAO for two instruments: TAIPAN [2] and MANIFEST [3,4,5] (Figure 4). These two projects involve developing a parallel fibre-positioner, using the Starbug technology, consisting of multiple autonomous miniature robots. The TAIPAN instrument will be placed on the UK Schmidt Telescope (UKST) at Siding Spring Observatory (SSO) in New South Wales, Australia and will perform as a spectroscopic facility tasked to observe some 500000 galaxies and 2000000 stars in a dedicated five year survey. The TAIPAN instrument is designed to use a complement of 300 Starbugs, with an initial deployment of the instrument, planned for the end of 2015, consisting of 150 Starbugs. TAIPAN will serve as a prototype for the MANIFEST fibre-positioning system for the Giant Magellan Telescope (GMT) [19]. The characteristics of these two instruments are summarised in Table 2.

Table 2: Proposed specifications of TAIPAN and MANIFEST

|  | TAIPAN | MANIFEST |
|---|---|---|
| Number of Starbugs | 150 (up to 300)  9 (guide bundles) | ~ 600 |
| Field of View | 6 degrees | 20 arcmins |
| Field ROC | 3.0 m | 3.275 m |
| Field diameter | 327 mm | 1250 mm |
| Field configuration time | < 5 minutes | < 8 minutes |
| Fibre configurations | 150 × single fibres | 420 × IFUs – 19 fibres (GMACS) |

|  | 9 × 1.5mm guide bundles | 1 × IFUs – 8400 fibres (GMACS) |
|  |  | 43/4 × Single fibres (G-CLEF) |
| Starbug exclusion radius (minimum separation between bugs) | 9 mm | 9 mm |
| Starbug patrol radius | 50 to 75 mm (science) | < 100 mm (science) |
|  | 10 mm (guide bundles) |  |
| Science Surveys and Data Sources (not definitive – estimated parameters as a guide only.) | TAIPAN – Galaxy: SDSS 14 < R < 16; exposures ~30 min x5=2.5 hrs; < 70% completeness<br><br>TAIPAN – Cosmology: SDSS 16 < R < 17; exposures ~30 min; < 70% completeness<br><br>FUNNELWEB – Stars: Tycho V < 11; exposures < 3 min; 100% completeness | - |

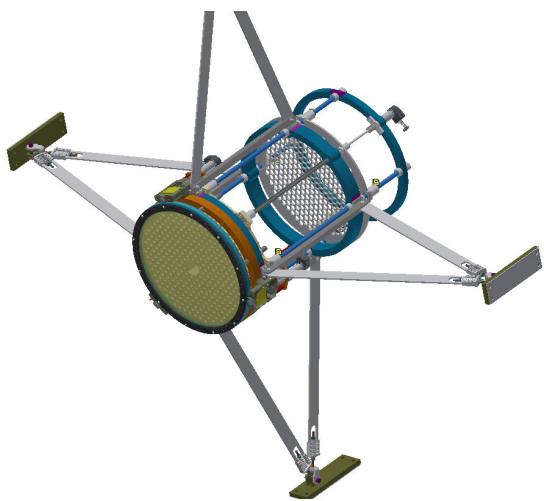

(a) TAIPAN – Schematic of focal plane assembly, a prototype for MANIFEST.

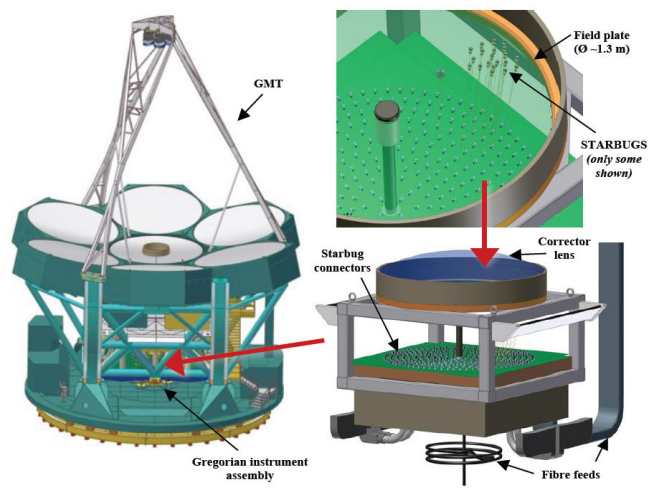

(b) MANIFEST – Located at the Gregorian instrument assembly and is moved in/out of the focal plane area.

Figure 4: Starbugs based instruments – TAIPAN and MANIFEST

## 2. SOFTWARE AND DATA MODELS

The goal of providing the fastest configurations times using Starbugs requires the optimal algorithms for the target allocation and routing modules. The ability to prototype and compare performance with a simulator tool is critical to the development. This is the case as a full complement of functional Starbugs have not yet been constructed. The algorithms must take into account the specifications of the Starbug and the instrument, such as the patrol radius, exclusion diameter, movement vectors, step sizes, velocity and rotation. These specifications are listed in Table 1 and Table 2.

The algorithms to be developed for the TAIPAN project should be easily adaptable to MANIFEST given the same Starbugs positioning technology. As Starbugs are unique when compared to other robotic positioners so the existing software base, such as 2dF Configure [20], cannot readily be used. Therefore, new algorithms are needed. A simulator tool is needed to visualise and analyse the algorithm performance as the TAIPAN software and hardware are currently in development. The three software sub-modules are listed in Table 3. The top-level TAIPAN software model diagram is shown in Figure 5 showing the location of the Allocator and Router.

Table 3: The software sub-modules for field allocation and routing tasks for TAIPAN.

|  | Task | Notes |
|---|---|---|
| Field Allocator | To allocate Starbugs to target objects (e.g. stars and galaxies). | Allocator's algorithms should find allocations that minimise the total field revisits, largest distances and path crossings to ease the Router's task. |
| Field Router | To find a path for each Starbug given its starting and ending position on the field plate. Takes allocation data as input. | Router's algorithm should find the shortest path for each Starbug that avoids collisions. |
| Simulator Tool | To read in the output of the router to provide a visualisation of the Starbugs moving along the field plane and generate some meaningful timing results. This module takes the place of hardware and software that is not yet developed. | Provide a realistic simulation of the Starbug performance to allow comparison between algorithms. |

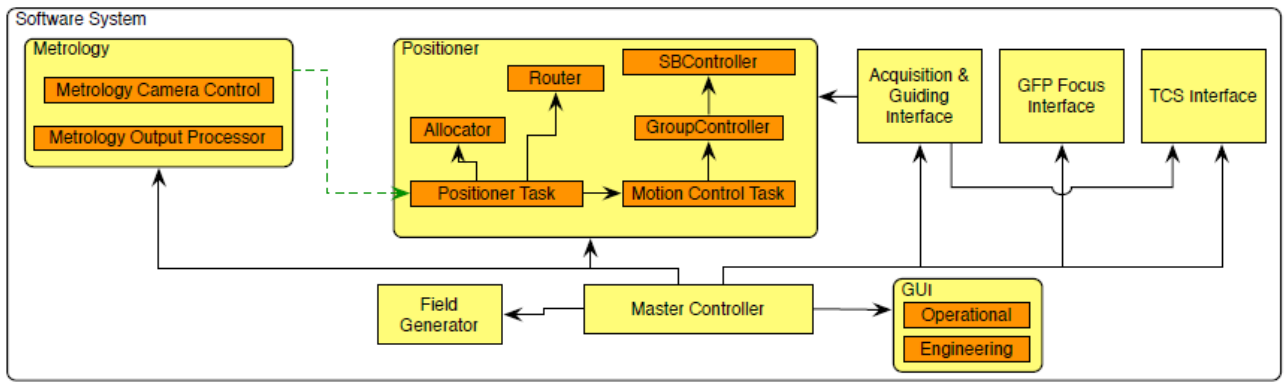

Figure 5: TAIPAN top-level software system showing the Allocator and Router as part of the Positioner Task. The Simulation Tool simulates the Metrology and Motion Control Task.

The data model for the allocation, routing and simulation tool modules is shown in Figure 6. The data is stored in XML format to allow a standard scheme to share the data between the various modules and external applications for viewing, e.g TOPCAT [22]. Having the data model specified provides the ability for multiple team members to develop the algorithms without running into significant compatibility issues.

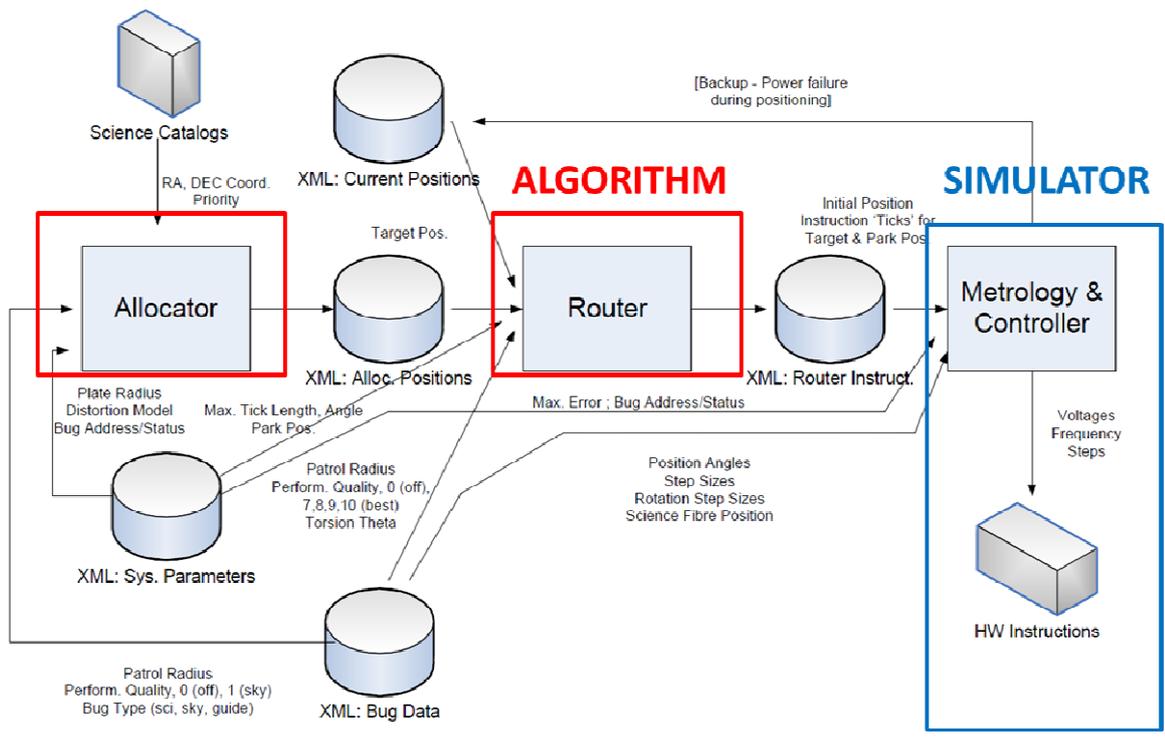

Figure 6: Data model for the Allocator, Router and Simulation Tool.

# 3. ALLOCATION ALGORITHMS

The Allocator sub-module assigns Starbugs to science targets and thereby determining each Starbug's next field plate position. It tries to optimise the allocation of all the Bugs in order to reach a high percentage of field completeness and to minimise both the number of crossings between them and the distance from the Bug to its next target. Contrary to the Router that will find, for each Starbug, the final collision-free path from its current to its required position, the Allocator only assigns the best target. This which will be — depending on the algorithm used — either the closest target, the one that minimises cross-over or the longest distance among all the bugs, or some further more complex algorithm.

The Allocator generates a list of XML files that will be used by the Router as its main input. Each file contains a VOTable which lists all the Bugs and their position on the focal plane for a particular field allocation. As each field can be composed of thousands of targets and only 150 Starbugs are available on the plate at any one time, the Allocator re-visits each field a number of times and employs a basic Closest Distance Algorithm (CDA-Basic) to allocate the positions, as illustrated in Figure 7. The CDA-Basic allocates, to each bug, the nearest available target. A hexa-polar arrangement for the initial ("park") positions of the Starbugs is found to provide most efficient arrangement of Starbugs for this algorithm. In order to generate the output files (one for each re-visit to a particular field on the sky) the Allocator uses as its input star or galaxy positions and priorities from science catalogues (Table 4). It also takes into consideration system and Starbug parameters such as the *Plate Radius* and Starbug *Patrol Radius* (Table 2).

In Figure 7 we see the CDA-Basic behaviour for the first 7 field re-visits; the field completeness is that of the ideal system (defined as exclusion radius of 0 mm and unbounded patrol radius). This allocation assumes the simple case of the park position as the initial Starbug position for each field revisit. For the $8^{th}$ and later field re-visits the slope of the completeness decreases (becomes flat) due to an increasing fraction of unallocated Starbugs. The unallocated Starbugs are a direct result of the clustering of the Galaxy targets, the limited patrol radius and the exclusion radius limit (i.e. Bug footprint area). After the closest targets have been allocated, the target distance begins to exceed the park position separation (24.1 mm) and the number of line-of-sight path intersections increases due to Starbugs moving further into adjacent neighbour regions. For the simplest case, the intersections should be minimised as they add delays to the route while the Starbug waits for another bug to cross and clear the line-of-sight path. At this point the number of intersections and maximum distance then decreases due to increased fractions of unallocated Starbugs. The field configuration time can be reduced by decreasing the maximum distance for each field re-visit, by ignoring the 5% of Starbugs having the largest travel distance, as shown in Figure 7 (c). This might be acceptable for the FUNNELWEB Survey (with required exposure times less than several minutes) to get the configuration times within the readout time of the CCD and hence eliminating downtime due to the fibre positioner operation.

Table 4: Source catalogues for testing the Allocator.

| Source Catalogue | Description |
| --- | --- |
| Galaxy-1 | 1,431 galaxies with approx. uniform priorities p1 (low) to p9 (high), mock sky fibres positions. |
| Star-1 | 5,916 stars with Galactic latitude 0 degrees, priorities (p9: mag. 9 $\leq$ mag. $\leq$ 11; p7 mag. = 12), mock sky fibre positions. |
| Star-2 | 427 stars with Galactic latitude 90 degrees, priorities (p9: mag. 9 $\leq$ mag. $\leq$ 11; p7 mag. = 12), mock sky fibre positions. |

Resolving the limitations of CDA-Basic, we have CDA-Advanced (see Table 5). In Figure 8, we see that the CDA-Advanced performs less well than the CDA-Basic as a result of the extra constraints placed on the algorithm. For a given field completeness the efficiency to allocate the highest priority P$\geq$9 is lower than including all the targets P$\geq$1. For later field re-visits, the bulk of the bugs have larger travel distances to targets. However, the CDA-Advanced does not have any line-of-sight crossings and each field begins from the previous field position. These optimisations improve the re-configuration times. Although CDA-Advanced has longer maximum Starbug travel distances than CDA-Basic, it does not require extra motion or 'traffic lights' that the Router produces to avoid crossings and collisions, and hence the re-configuration times are improved. CDA-Advanced should be capable of routing almost all fields (testing of this is in progress) unlike CDA-Basic.

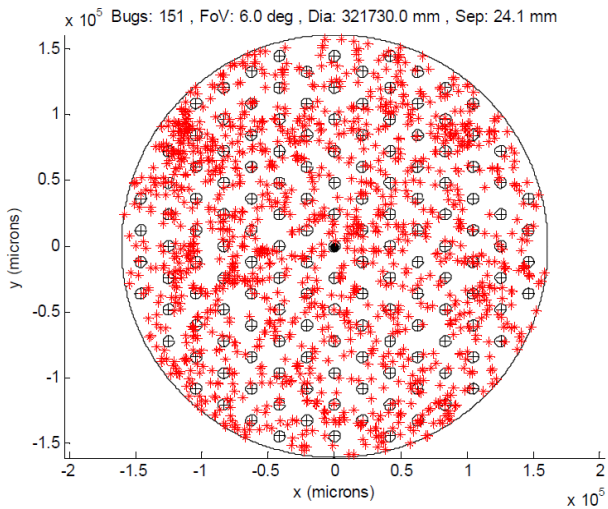

(a) Starbugs are represented by small crossed circles (at their park positions) and all the target positions for this field (1431 of them) are represented by the red stars.

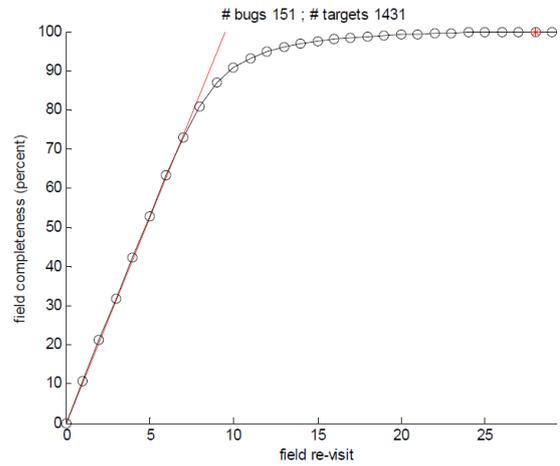

(b) The completeness percentage for the number of re-visits of this field. The red line is the case for the ideal system.

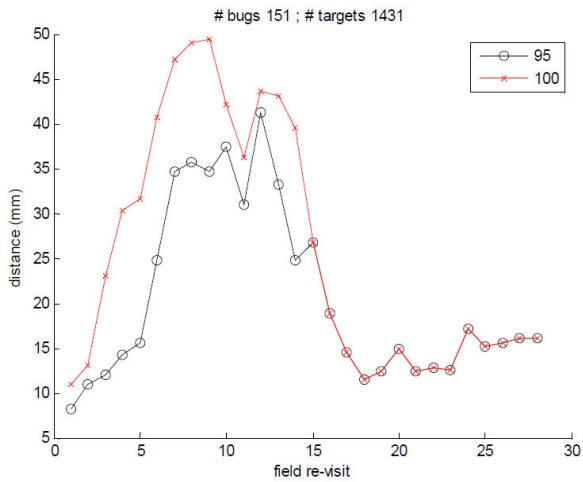

(c) The maximum distance a Starbug travels from park for a given field re-visit. The 95% distance represents the case that the top 5% of Starbugs having the largest distances are ignored (to improve field configuration times).

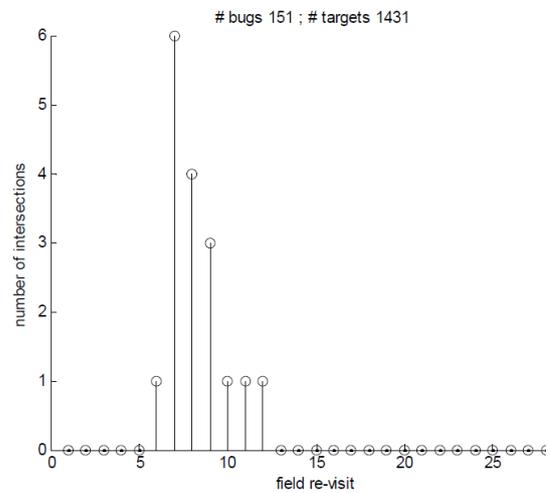

(d) The number of line-of-sight path crossings (intersections) for each field visit.

Figure 7: Illustrates the task of the allocator for Galaxy targets (CDA - Basic). Patrol Radius 50 mm. Source Catalogue: Galaxy-1. Each field-revisit begins from the Park position.

Table 5: Summary of the allocation algorithms for TAIPAN.

| Allocation Algorithm | Description | Negatives | Positives |
|---|---|---|---|
| Closest Distance Algorithm (CDA) - Basic | The Starbug with the closest target is allocated first. Then the next bug with the closest target. Continue until all bugs are allocated. Targets in the exclusion radius are removed. Each field revisit starts at the Park position. The movement sequence is therefore Park, Field 1, Park, Field 2, etc. | Some fields cannot be routed due to line-of-sight path contentions. Does not remove line-of-sight crossings that increase the complexity of the Router's task to include 'traffic lights' and/or A-star. Does not breakdown cluster regions resulting in efficiency losses at larger field revisits due to unallocated Bugs. Does not consider the use of target priorities or Sky Fibres. Inefficiency due to starting from the Park position for each revisit. | Performance similar to that of an ideal positioner for completeness less than 70%. The first set of field re-visits has the lowest distances and hence lowest configuration times. The 95% distance shows significant reduction gains in allocation time. A foundation for further algorithm development. |
| Closest Distance Algorithm (CDA) - Advanced | Similar to CDA-Basic except in handling of the negatives: line-of-sight crossings are removed, and target priorities and sky positions are included. The algorithm also attempts to break cluster regions down into a series of sparser fields. Unlike CDA-Basic, each field revisit starts at the previous field position. | Poorer performance (completeness and distance) than CDA-Basic. | An implied assumption that all fields can be routed successfully. Improves the re-configuration times. |

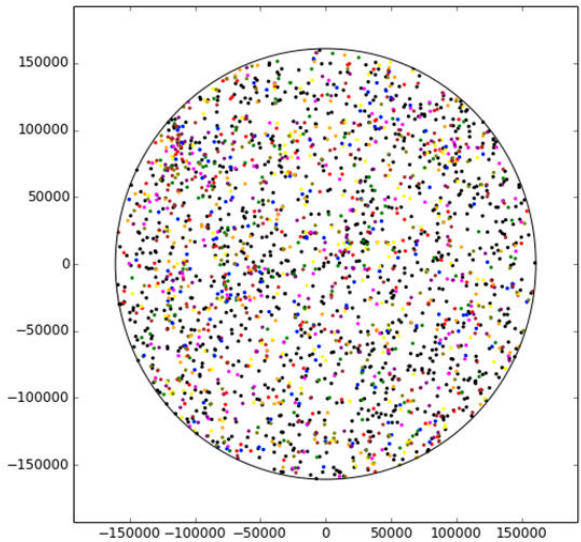

Positions of target galaxies and skies (black), in units of microns. Priority: 9 (red), 8 (yellow), 7 (green), 6 (blue), 5 (magenta), 4 (orange), 3 (olive), 2 (purple), 1 (brown).

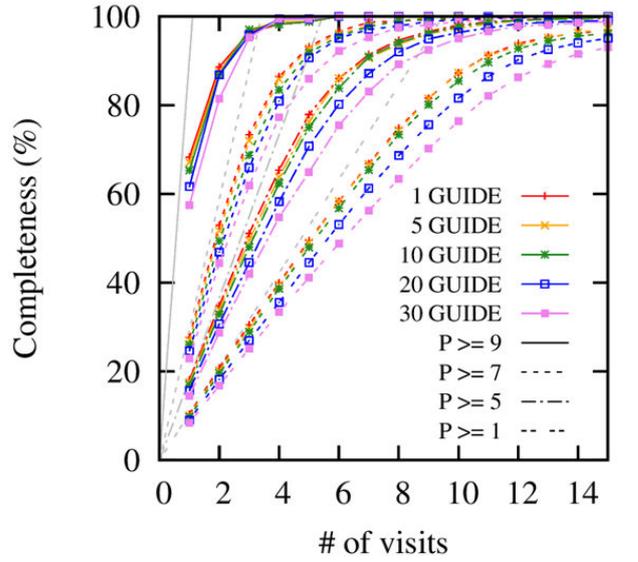

Target completeness over number of visits, for targets above various priority thresholds, with various choices of minimum number of sky fibers (GUIDE). Gray lines correspond to the ideal case, where all 151 Starbugs successfully allocate new targets at each visit.

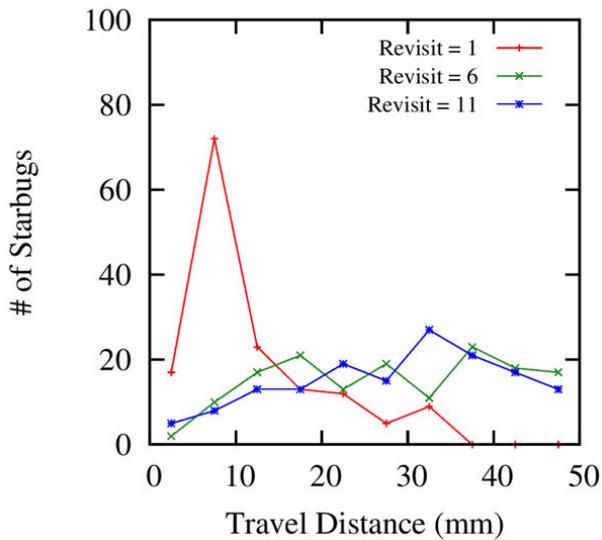

Distribution of Starbug travel distance for given visits.

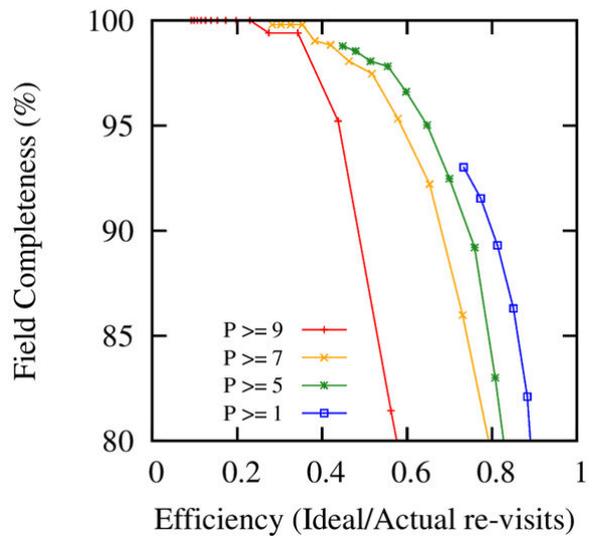

Efficiency vs. field completeness for various priority thresholds, with the minimum number of sky fibres is 30.

Figure 8: Illustrates the task of the allocator for Galaxy targets (CDA-Advanced). Patrol Radius=50 mm. Source Catalogue: Galaxy-1. Each field-revisit begins form the previous field position.

Figure 9 and Figure 10 show the results using CDA-Advanced from two source catalogues on stars, with different number of stars (Star-1 and 2). While the final field completeness of Star-1/2 is similar to Galaxy-1 (~93%), the efficiency is enhanced, mainly due to the uniform spatial distribution of stars compared to the galaxies. Also, the efficiency decreases as the number of targets increases.

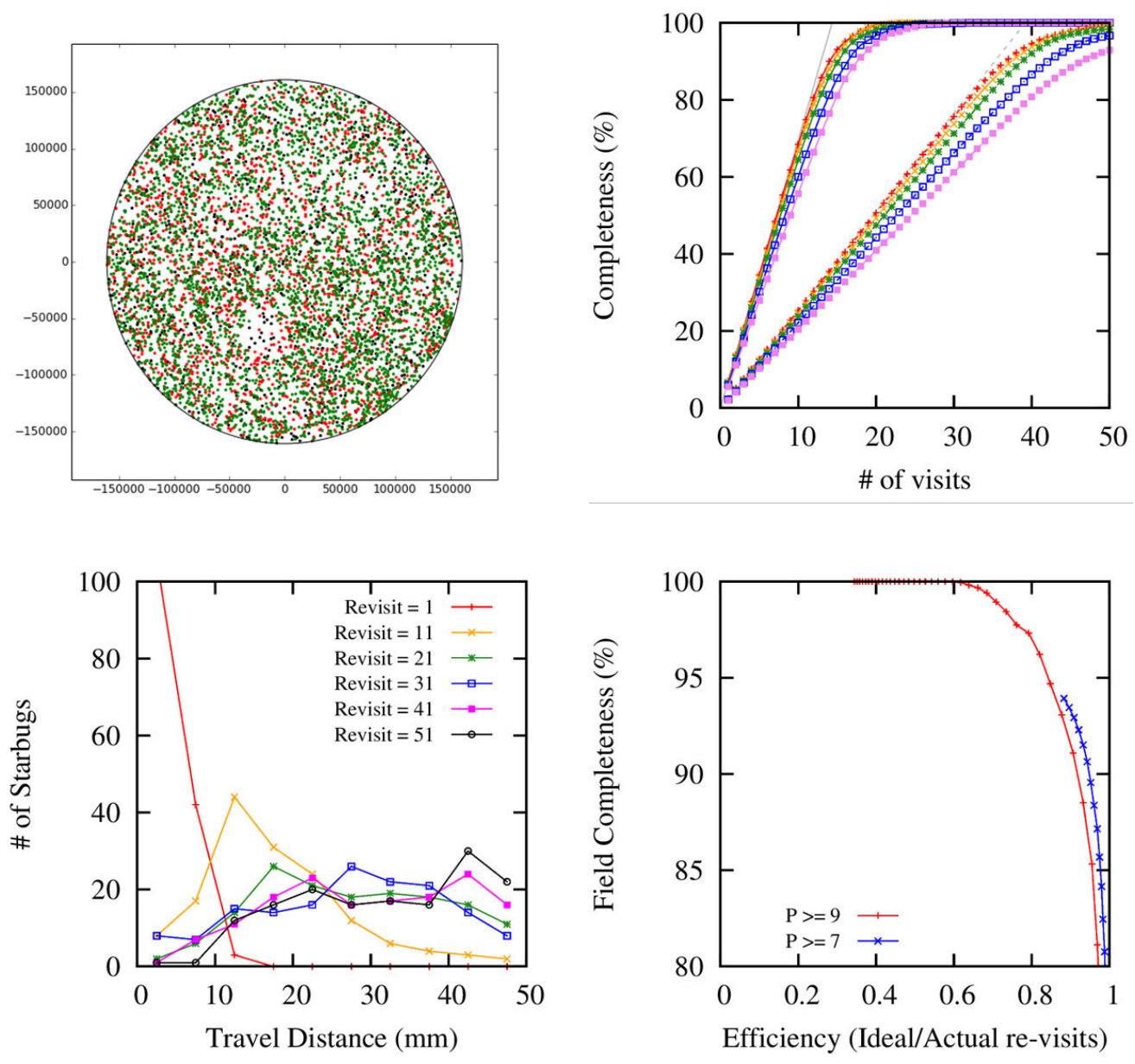

Figure 9: Illustrates the task of the Allocator for Star targets (CDA-Advance). Patrol Radius 50mm. Source Catalogue: Star-1. Each field-revisit begins form the previous field position.

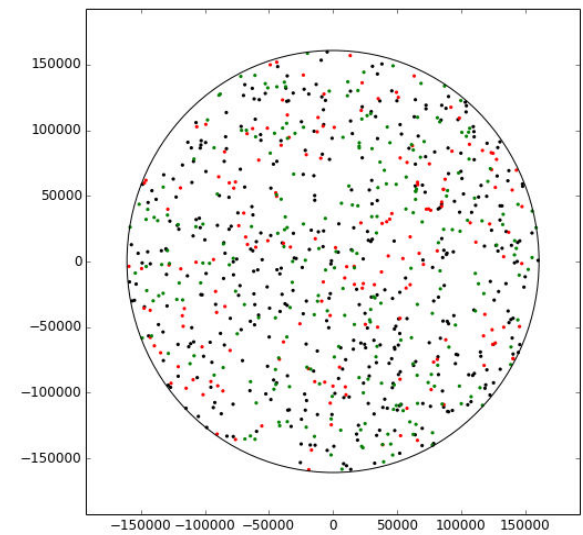
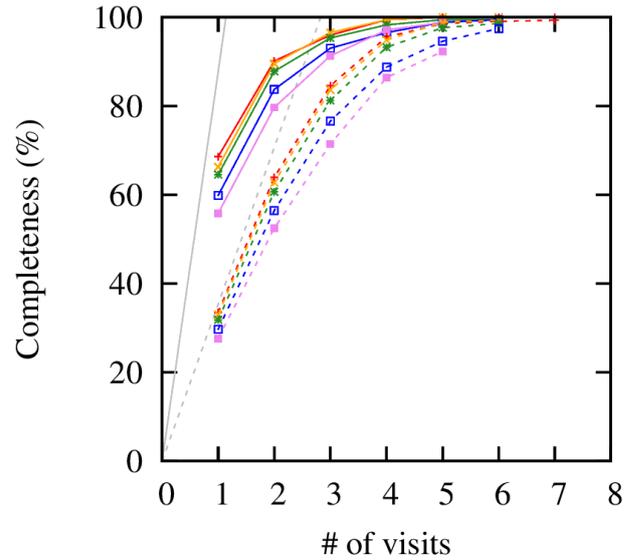
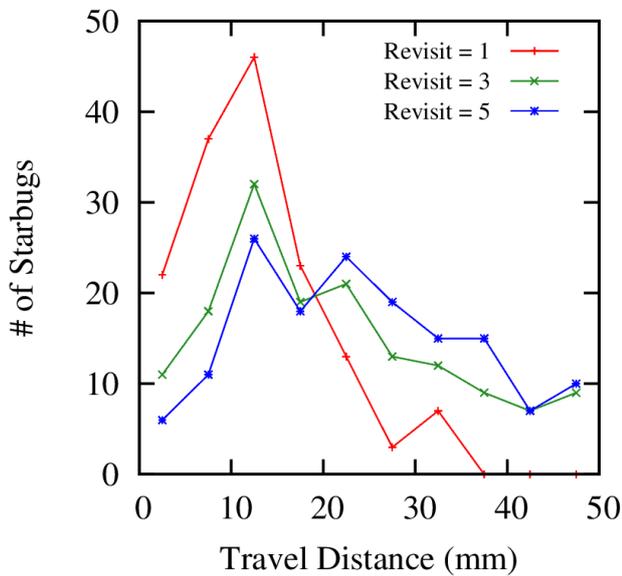
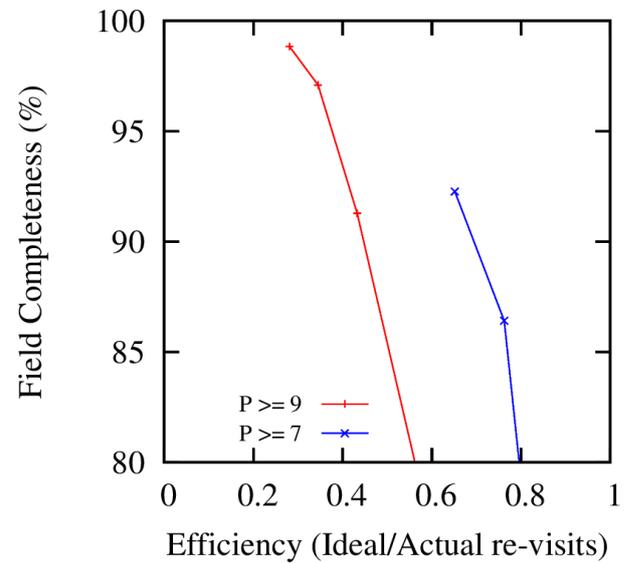

Figure 10: Illustrates the task of the Allocator for star targets (CDA-Advanced). Patrol Radius 50mm. Source Catalogue: Star-2. Each field-revisit begins form the previous field position.

## 4. ROUTING ALGORITHMS

The Router sub-module has the challenging task to find a collision-free path for each Starbug as directed by the allocation file. In addition, it should also find two other collision-free paths for each one of these files. One to route the bugs from their current positions to their Park positions and one to reach their targets from the Park positions (see Figure 11). As illustrated in the figure the Router takes as input a list of allocation files (a minimum of two, which are provided by the Allocator), the system parameters (for the park positions), the Starbug data (for the Patrol Radius) and a backup file (provided by the Metrology and Controller sub-modules) storing the current position of the bugs in case of a power failure during positioning. The Router's output is a list of XML files in which the three sets of paths described above are converted into movement instructions. Figure 11 shows the structure of these files. There are four VOTables in each file: one for the initial position and one for each list of paths that the Router has computed and turned into instructions. These instructions will be used by the Controller sub-module to actually move the Bugs.

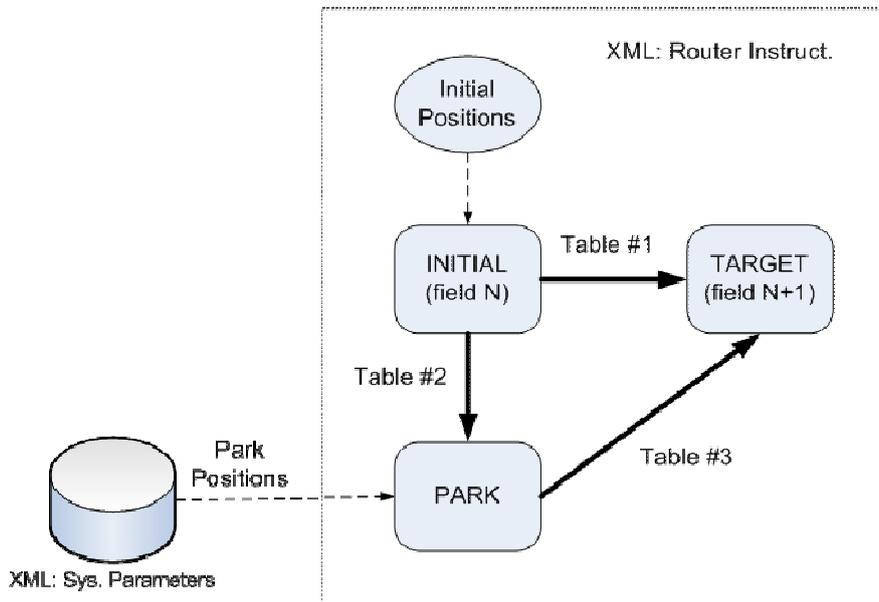

Figure 11: The Router's output XML file is composed of four tables. One containing the initial positions of all the Starbugs, one with the instructions to move the bugs from their initial positions to their target positions (table # 1), one to move them from their initial positions to their park positions (table # 2) and finally one to move them from their park positions to their target positions (table # 3). The park position of each Starbug is stored in the *XML: System Parameters* file which is an input of the Router.

It is important to understand here why we need these two other paths. Generally the astronomers that will use TAIPAN will select, during the day, a set of fields they want to observe for the coming night and generate first the allocation files and then the Router's instructions. It is likely, however, that during the course of the observation the order of the files may need to be modified (i.e. poor weather or a change of mind). The computation of these two extra paths are therefore more than vital. Indeed, they allow the Controller to send the bugs back to their park position from any observing position using table #2 of Figure 11 and then to change the target by using table#3 of the new field to be observed. There's then no need to recompute all the paths which will save precious amount of telescope time.

An important concept introduced with these files is the concept of a 'Tick' (see Figure 12). A tick delimits the motions of the bugs and with that plays the role of the Router algorithm's structure. A Starbug move instruction is composed of the Tick number, Bug ID, the final position of the motion (i.e. x, y and theta), and the actual instructions to reach this final position (i.e. dX, dY and θ). A Tick can therefore be seen as a set of synchronous bug movements from a known initial state to a known final state with no path collisions. All Bug movements must be completed for Tick N before executing Tick N+1. A Bug's total path may consist of many Ticks from its current position to its destination. A Bug can also be kept in a 'waiting' state if it is not assigned to move during a given Tick.

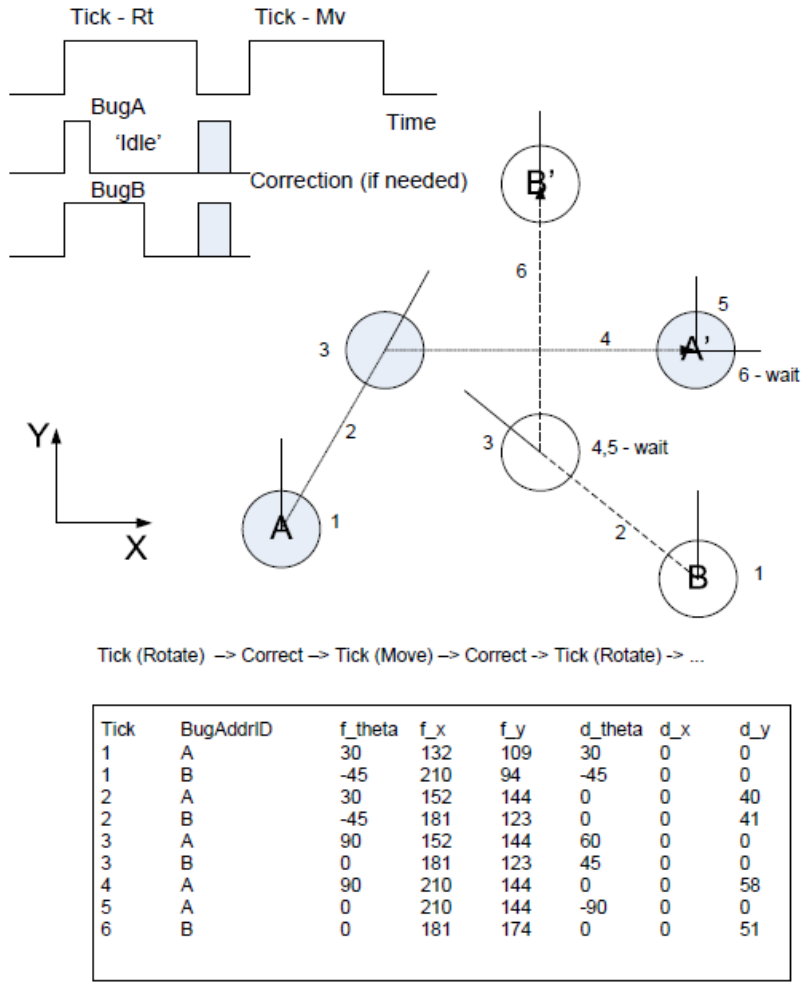

Figure 12: Illustration of the Tick Concept.

The design of the Router algorithms has been done in stages to reach a more complex and efficient algorithm. The goal is to breakdown the complexity of the task at hand into smaller and simpler ones to measure and compare performance. Three Router algorithms have been developed, starting from Simple Vector, Traffic Light then finally Traffic Light and Cooperative A*. A summary description of each Router Algorithm is listed in Table 6. A detailed description of the Router algorithms can be found in Satorre (2014) [21].

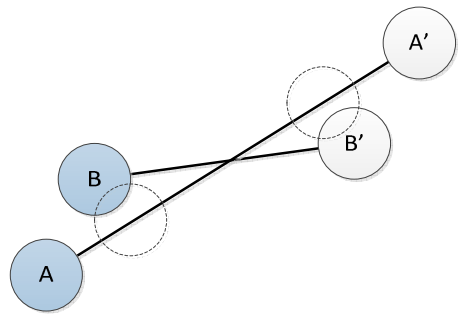

Figure 13: An unavoidable collision for line-of-sight path crossing (A to A' and B to B') that Traffic Light fails (see Table 6).

Table 6: Description of the three Router algorithms.

| Router Algorithm | Description |
|---|---|
| Simple Vector | Simple algorithm that has a non-optimal solution in terms of configuration time (similar to a sequential positioner) but without any collisions so that it can be used as a base model for comparison with the other algorithms. Starbugs that have no path crossing are moved first to destination. A bug in the crossing pair is also moved first. Then the remainder of bugs are moved to destination. Note that some fields are not routable but this problem can be solved with the Allocator (CDA-Advanced). |
| Traffic Light | This principally gives the bug the ability to stop and wait somewhere along its path in order to avoid collisions. However, there is the case of the unavoidable collision (as both, the Start and End positions of B are too close to A's path a collision will occur no matter which Bug moves, see Figure 13). As with the Simple Vector algorithm, some fields are not routable but can be solved with the Allocator (CDA-Advanced). |
| Traffic Light and Cooperative A* | Bugs that cannot be routed with the Simple Vector and Traffic Light algorithms (see Figure 13) need a more complex solution that can generate a path avoiding collision by making detours. A modified version of the CA* algorithm has been implemented for this purpose. |

## 5. SIMULATOR AND RESULTS

The Starbug Control System closed loop ensures the correct positions of each Starbug while it is moving. The Metrology software module has the function to provide to the Positioner (Controller) the Starbug position information. This will keep track of where each Starbug is at any given time. As part of the Starbug Control closed loop, the Starbug Motion Controller will first use the Router's instruction to drive the Bugs and then use this information to check that the Bugs follow these instructions correctly. If they don't the Controller can modify the behaviour of each bug by adding new correction instructions.

A Simulator Tool was designed early in the project to somewhat mimic the basic functionality of the Metrology and Positioner modules. The Simulator Tool is valuable in testing the output of the Router, to measure the time taken for each field configuration and to check for collisions. This has been implemented using *Qt Creator 2.4.1* based on *Qt 4.8.0* on a Linux machine running *Ubuntu*.

The user interface of the Simulation Tool is shown in Figure 14. The main window (GUI) of the tool is divided into six parts. On the left there are three Dock Widgets, each allowing the user to set different parameters of the simulation (e.g. the input files, the algorithm to use, or the voltage and frequency that will determine the speed of the Bugs). The Central Widget (on the right) contains the current simulation. On the top of the window the Menu Bar allows the user to run the Router's program and to launch the simulation. Finally on the bottom is the Status Bar that displays the running time of the simulation.

The inputs of the Simulator Tool are the XML files generated by the Router (see Figure 11 for details). It currently doesn't take into account the System Parameters or the Bug Data (Figure 6). From the Allocation Dock Widget the user can select the files that the Router will use as input and then directly run it from the Menu Bar. The XML files can then be selected through the Algorithm Dock (the Router will generate a set of files for the three algorithms developed), choose which path to simulate (one of the three paths illustrated in the figure) and start the simulation from Menu Bar.

The output of the Simulator Tool is in the form of a simple CSV file that can processed using an interactive graphical viewer such as TOPCAT [22] or MATLAB. This allows the output of the simulations to be easily compared.

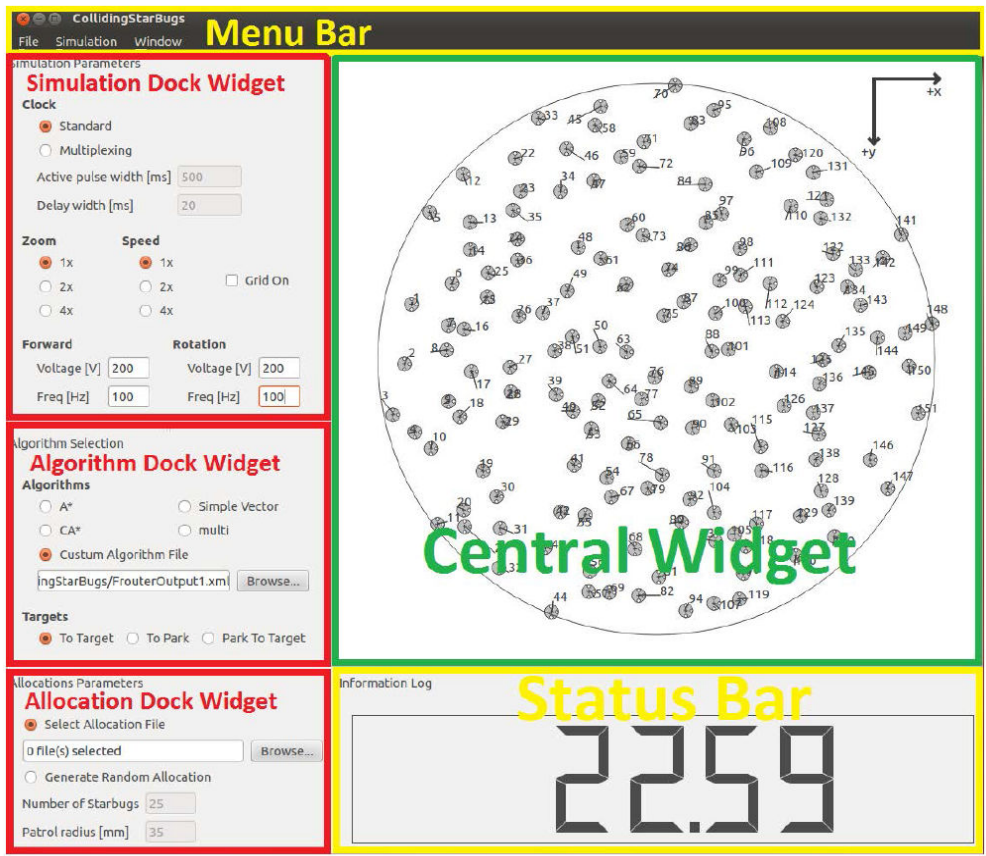

Figure 14: The Simulator's Graphical User Interface which is separated into six parts. Three Dock Widgets, one Central Widget, one Menu Bar and one Status Bar.

Representative results comparing the Router Algorithms are shown in Figure 15 for each allocation file (complex set). The following can be stated:

- The median values of Simple Vector (SV) are smaller than the ones of Traffic Light (TL) and Final (F, Traffic Light and Cooperative A*) for the first three fields of the complex allocation set. This has, for now, no explanation and should be studied in depth. The improvement of the configuration time for TL compared to SV is even more substantial than for the previous set, as SV reaches very long configuration times. The maximum reduction is 71% for field number 3.

- For this set of allocation files, the collision results aren't good at all. For all the fields, collisions occur for all the algorithms. F compared to SV has always fewer or the same number of collisions, but generates an extra collision in the last field compared to TL. This extra collision as well as the unexpected high number of collisions in fields 3, 4 and 5 show a failure in the implementation of the CA* algorithm. TL repeats its weird behaviour already noticed in the previous set of allocation files, showing more collision than SV (fields 2 and 3).

- Once again we can see that by applying the 95% filter, significant savings are made. Further study will be carried out in order to determine if this case also reduces collisions.

- The configuration time of F is maximal for field 5 with 80.21 s and 10 collisions. Using the 95% filter this can be reduced to 50.48 s, which corresponds to a reduction of 37%.

- The results for the complex allocation set have revealed some problematic failures in the implementation side of the CA* algorithm and have also confirmed the presence of errors in the TL.
- The work on these algorithms is continuing so that found issues are resolved.

Figure 16 shows some particular behaviours among the Starbugs and illustrates some of the representative results shown in Figure 15. The timings for each Starbug are shown: the red points represent the Rotating Times, the blue ones the Moving Times and the green ones the Waiting Times.

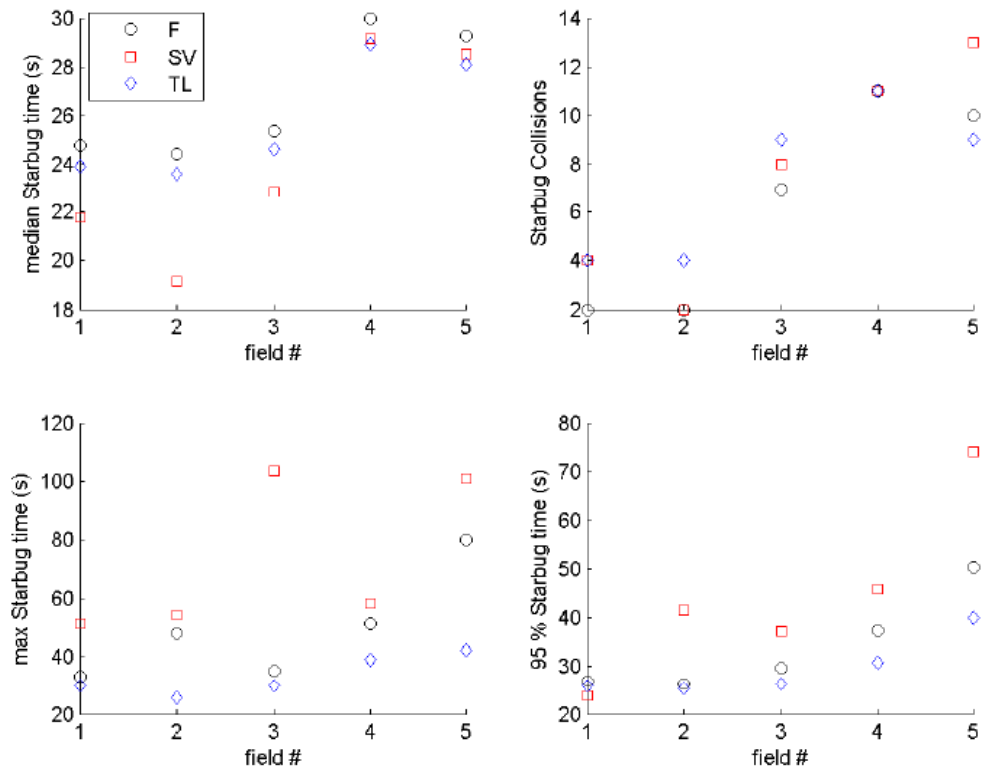

Figure 15: These plots show for each allocation file (complex set) and for each algorithm, the representative results of the simulation, with the number of collisions (upper left), the configuration time (i.e. the longer total time among the bugs, lower right), the median among all the total times of the bugs (upper right) and the time required for 95% of the bugs to reach their final position (lower left).

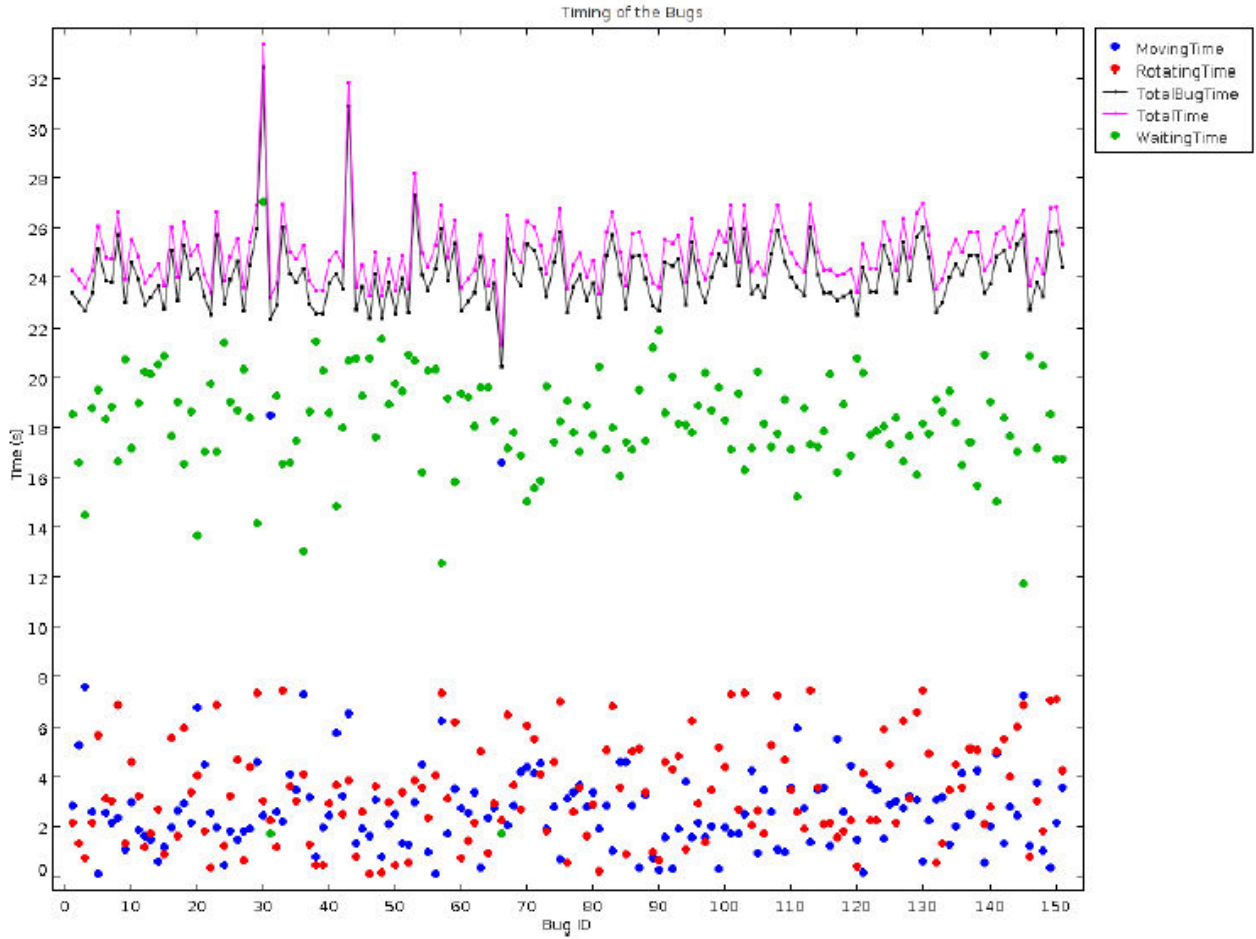

Figure 16: This figure illustrates the timing of the Starbugs for the Complex Allocation set. Algorithm used: Traffic Light and CA* Fields: 1 to 2; Maximum Time: 33.3 s; Collisions: 2

## 6.  CONCLUSIONS

The field configuration algorithm for Starbugs is critical to achieving the fast reconfiguration time goals desired for TAIPAN instrument. The field configuration should be generalised for compatibility with the MANIFEST instrument. Starbugs is a new positioner and as such provides unique challenges. We have identified the key software modules and have prototyped two Allocator algorithms (CDA-Basic and CDA-Advanced) and three Router algorithms (Simple Vector; Traffic Light; Traffic Light and Cooperative A*).

We have learnt that the Allocator can ease the task of the Router with minimizing path crossing and reducing path distances. The CDA-Basic works well if the completeness of the Survey is below 70%, otherwise the number of path crossings increase, as well as the fraction of unallocated bugs due to target clustering. Configuration times can be further improved (perhaps within CCD readout) by ignoring Starbugs in the top 5% of distance to travel between fields – this has substantial gains for the FUNNELWEB Survey where field exposures less than three minutes are expected.

We have found that the Router is a complex module and needs prototyping in several stages of increasing complexity. Three methods were presented for the Router. Simple Vector approach is the slowest (final Bug to destination) and does not optimally handle the more complex fields where path crossings are required. The Router needs to include wait states, or Traffic Lights to efficiently handle path crossings (minimise Bug total waiting time). However, including Traffic Lights does not solve the problem for the case of unavoidable collision (Figure 13). For this case the modified Cooperative A-star algorithm is used but field configuration times are increased compared to Traffic Light. In principle, the Traffic Light and modified Cooperative A-star algorithm should be capable of successfully routing most fields. The modified CA* algorithm needs further development to resolve implementation issues, but is our best candidate to date.

The Simulation Tool has been a useful diagnostic to prototype our algorithms. The output allows visualisation and performance comparison to support algorithm development. By prototyping and assessing performance with the Simulation Tool we have gained insight to the challenges ahead for 150-Bug TAIPAN instrument on the UKST.